\documentclass[twoside,11pt]{article}

\usepackage{blindtext}

\usepackage{jmlr2e}
\usepackage{comment}

\usepackage{lastpage}
\jmlrheading{23}{2026}{1-\pageref{LastPage}}{TBA; Revised TBA}{TBA}{21-0000}{Prospect Collaboration}

\ShortHeadings{GAPE}{Adriamirado et al.}
\firstpageno{1}

\let\oldcite\cite
\renewcommand{\cite}[1]{\textit{\oldcite{#1}}}

\usepackage[left]{lineno}  

\usepackage{listings}
\usepackage{xcolor}

\begin{document}


\title{New Deep Learning Data Analysis Method for PROSPECT using GAPE: Genetic Algorithm Powered Evolution}

\author{
{\normalsize \noindent {\bf M. Adriamirado} \and {\bf O. Benevides Rodrigues} \and {\bf A. Irani} \and \\ {\bf B. R. Littlejohn} \and  {\bf F. Machado}} \\
\addr Illinois Institute of Technology, Chicago, IL 60616, USA \\
{\bf A. B. Balantekin} \\
\addr University of Wisconsin, Madison, WI 53706, USA \\
{\bf C. Bass} \\
\addr Le Moyne College, Syracuse, NY 13214, USA \\
{\bf E. P. Bernard} \and {\bf N. S. Bowden} \and {\bf T. Classen} \and {\bf S. Ghosh} \and {\bf M. P. Mendenhall} \and {\bf C. Roca} \and {\bf X. Zhang} \\
\addr Nuclear and Chemical Sciences Division, Lawrence Livermore National Laboratory, Livermore, CA 94550, USA \\
{\bf C. D. Bryan} \and {\bf A. J. Conant} \and {\bf G. Deichert} \and {\bf M. Fuller} \\
\addr High Flux Isotope Reactor, Oak Ridge National Laboratory, Oak Ridge, TN 37831, USA \\
{\bf N. Craft} \and {\bf M. J. Dolinski} \and {\bf C. E. Lane} \and {\bf A. Lozano Sanchez} \and {\bf R. Neilson} \\
\addr Drexel University, Philadelphia, PA 19104-2875, USA \\
{\bf A. Delgado} \and {\bf P. E. Mueller} \\
\addr Physics Division, Oak Ridge National Laboratory, Oak Ridge, TN 37831, USA \\
{\bf A. Erickson} \\
\addr Georgia Institute of Technology, Atlanta, GA 30332, USA \\
{\bf A. Galindo-Uribarri} \and {\bf B. Heffron} \and {\bf D. Venegas-Vargas} \\
\addr Physics Division, Oak Ridge National Laboratory, Oak Ridge, TN 37831, USA and \\
University of Tennessee Knoxville, Knoxville, TN 37916, USA \\
{\bf S. Gokhale} \and {\bf S. Hans} \and {\bf R. Rosero} \and {\bf M. Yeh} \\
\addr Chemistry Division, Brookhaven National Laboratory, Upton, NY 11973, USA \\
{\bf C. Grant} \\
\addr Boston University, Boston, MA 02215, USA \\
{\bf A. B. Hansell} \\
\addr Susquehanna University, Selinsgrove, PA 17870, USA \\
{\bf T. E. Haugen} \and {\bf H. P. Mumm} \\
\addr National Institute of Standards and Technology, Gaithersburg, MD 20899, USA \\
{\bf K. M. Heeger} \and {\bf J. Wilhelmi} \\
\addr Yale University, New Haven, CT 06520, USA \\
{\bf J. Koblanski} \and {\bf J. Maricic} \and {\bf A. M. Meyer} \and {\bf R. Milincic} \\
\addr University of Hawaii, Honolulu, HI 96822, USA
}

\makeatletter
\def\@editor{}
\def\@starteditor{}
\def\@endeditor{}
\makeatother

\maketitle

\begin{center}
\small Correspondence: \texttt{prospect.collaboration@gmail.com}
\end{center}

\newpage

\begin{abstract}
We propose a genetic algorithm powered evolution (GAPE) method to create deep learning solutions for energy and position estimation for reactor antineutrino interactions in the Precision Reactor Oscillation and Spectrum Experiment (PROSPECT) at the highly enriched High Flux Isotope Reactor (HFIR) at Oak Ridge National Laboratory.  
We also apply GAPE to create classification models to distinguish signatures of inverse beta decay (IBD) interactions of reactor antineutrinos from common background types. The GAPE method can also be adopted for optimization of other types of problems that utilize machine learning (ML) models for particle physics applications.  
When applied in the PROSPECT context, we find that the models selected by GAPE can, in some cases, outperform the traditional models previously used for PROSPECT data analysis.  
In particular, when benchmarked against conventional PROSPECT neutrino identification pathways using the same underlying information, the classifier offers the promise of improving the signal-to-background ratio by nearly 2.8 times.  
Performance biases uncovered during initial IBD classifier validation were primarily caused by differences in time-dependent response between background and signal training datasets.  
Biases were effectively mitigated through a data-period-specific training regimen, offering a pathway towards realizing an unbiased IBD signal classifier for future reactor neutrino datasets.  

\end{abstract}


\section{Introduction}

Machine learning techniques are now widely used in the particle physics community for a variety of purposes, from particle tracking and reconstruction~\cite{Newby9026} to full waveform analysis in Time of Flight (TOF) experiments~\cite{jocher2018wave}. Recent neutrino experiments that have utilized ML methods include DUNE~\cite{Abi_2020}, MicroBoone~\cite{Acciarri_2017}, IceCube~\cite{Abbasi_2021}, and PROSPECT~\cite{andriamirado2025machinelearningsingleendedevent}. A 2022 survey of ML methods in particle physics experiments was published by Nature~\cite{Radovic2022}. 

Our paper introduces a genetically enhanced deep learning approach for reconstructing the physical attributes of MeV-energy neutrinos produced inside the High Flux Isotope Reactor (HFIR) at Oak Ridge National Laboratory (ORNL) and detected inside the PROSPECT (Precision Reactor Oscillation and SPECTrum) particle physics detector.  In particular, we will determine how this deep learning approach, referred to as GAPE, compares to PROSPECT’s traditional methods in separating true neutrino interactions from common backgrounds and in reconstructing the true energy and location of the interacting neutrinos.  Background rejection is an important aspect of PROSPECT performance, since the detector seeks to measure weakly-interacting neutrinos in the high-radiation environment of an on-surface nuclear reactor facility.  Neutrino energy reconstruction is also important for PROSPECT measurements, since a primary deliverable of the experiment is to perform measurements of the aggregate energy spectrum of HFIR’s neutrinos.                                                                                                                                                                                                                                                                                                                                                                                                                                                                                                                                                                                                                                                                                                                                                                                                                                                                                                                                                                                                                                                                                                                                                                                                                                                                                                                                                                                                                          There has been sizeable interest recently in the study of reactor antineutrino energy spectra, as there have been observed disagreements \cite{Almaz_n_2020,2021,An_2012,PhysRevLett.108.131801,PhysRevLett.108.191802}  with predictions based on beta-spectrum conversion \cite{Hayen_2019, Huber_2011} and calculations from nuclear databases \cite{Estienne_2019}. These discrepancies report excesses of detected interactions of neutrinos in the region between 5-7 MeV.

\begin{figure}[!htbp]\centering
\includegraphics[width=1\linewidth]{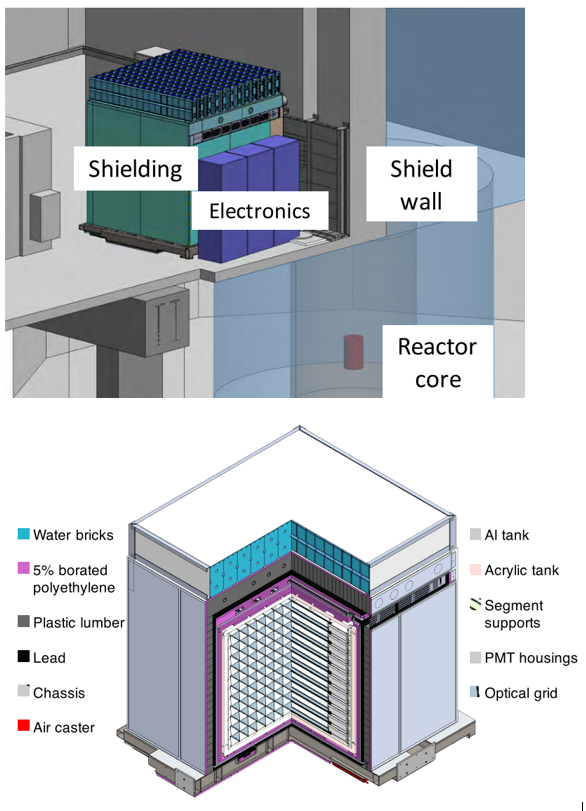}
\caption{(top) Diagram \cite{2019} of the layout for the experiment. Detector center to the reactor core center is approximately 7.84m. (Bottom) Diagram ~\cite{2019} showing cross-section of the PROSPECT detector. }
\label{f1}
\end{figure}

PROSPECT was designed to make precision measurements of the antineutrino spectrum and  probe eV scale massive sterile neutrinos by searching for neutrino oscillations over meter-long baselines from a highly-enriched uranium reactor~\cite{2019}.  The 4-ton antineutrino detector target, shown in Figure \ref{f1}, consists of a 14 by 11 grid of long, rectangular segments, filled with a ${}^{6}_{}\mathrm{Li}^{}_{}$-doped liquid scintillator to detect antineutrinos through inverse beta decay (IBD) while efficiently rejecting backgrounds. Each segment is 1.176 m in length and has a square cross-sectional area of 0.145 m × 0.145 m, with  a photomultiplier tube (PMT) in an acrylic housing on both ends of each segment. 240 Hamamatsu R6594 PMTs and 68 ET 9372KB PMTs were used in the experiment. The detector also contains gamma ray and neutron shielding, along with detector movement elements (in practice the detector was never moved). For a comprehensive review of the many elements of the PROSPECT shielding see~\cite{2021}.  

PROSPECT collected physics-quality data for 96 days when the HFIR reactor was on and 73 days when the reactor was off, with all data collected at an average baseline distance of 7.84 m from the center of the 85 MW HFIR core. Using traditional analysis techniques, PROSPECT detected a total of 115,852 IBD-like events in the reactor-on data set \cite{2021}, with 28,358 ± 18 being contributed by reactor-related, non-neutrino backgrounds, a category that we will describe later in greater detail and will hereafter refer to as ‘accidental’ backgrounds.A total of 30,568 IBD-like signals detected in the reactor-off dataset were contributed overwhelmingly by cosmic ray interactions inside PROSPECT, a background contribution that is largely consistent between reactor-on and reactor-off periods. Following live-time and atmospheric
pressure scalings and subsequent subtraction of these backgrounds, we are left with a total of 50,560 ± 406 interactions of electron antineutrinos \cite{2021}. The majority come from beta decays of ${}^{235}_{}\mathrm{U}^{}_{}$ fission products but approximately 1$\%$ come from non-equilibrium isotopes and non-fuel sourced antineutrinos from ${}^{28}_{}\mathrm{Al}^{}_{}$ and ${}^{6}_{}\mathrm{He}^{}_{}$.  For more information on the detector design, experiment layout, experimental results, as well as detailed information on calibration efforts, see ~\cite{2019} and ~\cite{2021}.  

PROSPECT detects HFIR-produced neutrinos by sensing the products of their interaction with a proton via IBD. In IBD events, the antineutrino interacts with a proton in the detector to produce a positron and a neutron. The positron carries most of the antineutrino energy and quickly annihilates with an electron, producing gamma rays, while the neutron thermalizes and then most often captures on a $^{6}$Li nucleus (approximately 75$\%$ of the time), which decays into a triton and an alpha particle, depositing energy.  Both of these time-correlated events emit visible photons, which are converted into photoelectrons (PEs) and counted by the PMTs at the end of each segment ~\cite{2021}.  

Thus, IBD candidates are selected by searching for two events close to each other in time and position, with the earlier of these events, referred to as the "prompt" event, having characteristics of positron annihilation, and the later event (referred to as the "delayed" event), having characteristics of a neutron capture. Important characteristics used here are the total energy measured in each event, as well as the PMT pulse shape discrimination (PSD) that measures the relative intensities of fast and slow scintillation light at each PMT. The PSD is distinctly different for neutron captures as it is for electromagnetic signatures. A more detailed discussion of the IBD selection cuts is outlined in section \ref{Sim Perf}.  

The application of these IBD signal analysis cuts to PROSPECT data during reactor-on periods will result in selection of a combination of true neutrino interactions as well as backgrounds from two primary categories.  First, random yet spatially and temporally coincident events from physically independent interactions, such as a reactor gamma scatter and a cosmic neutron capture, can mimic a truly correlated prompt and delayed IBD signatures.  This background category greatly increases during reactor-on periods due to the high reactor gamma fields surrounding PROSPECT.  The second dominant background type arises from truly time-correlated signatures of cosmic neutrons, such as a prompt neutron inelastic scatter followed by a delayed capture of the same neutron.  This background category remains largely unchanged between reactor-on and reactor-off periods.  

Position reconstruction in the context of IBD selection refers to the determination of the initial segment of interaction (SOI) for a particular event. The method for determining primary segment number used in the traditional analysis relies on selecting the segment with the highest-energy contained pulse of photon arrivals \cite{2021}.  For energy reconstruction, the traditional method employed by PROSPECT fits a maximum-likelihood function of the total photoelectrons collected (PEs) and the reconstructed position along the length of the segment to calculate the energy deposited during the event.

A fraction of PMTs malfunctioned due to current instabilities during the detector run. Additionally, a gradual decay in scintillation was observed due to interactions between the scintillator and detector materials and a slow precipitation of lithium from the scintillator reduced the neutron capture efficiency with time. The approach in this work mirrors the previous PROSPECT analysis ~\cite{2021} in which a single detector configuration was used in the final analysis, which excluded information from any 'dead' segments where one or both PMTs failed at any time during the detector operation. A recent PROSPECT effort to measure the antineutrino spectrum\cite{adriamirado2023final} yielded a dataset with greater statistical power and utilized several different detector configurations to more optimally combat changing detector conditions.

\section{GAPE Method}

Genetic algorithms (GA) \cite{GA} are useful for finding high-grade solutions to problems where gradient descent is not a viable option. They provide a flexible and automated framework for jointly optimizing model architecture, hyper-parameters, and feature selection in a high-dimensional space. Deep learning problems fall into this category, as often alterations in the optimum of one parameter may change the optimum of the others. 

In this section, we discuss how GAPE was used in selecting the solutions to three vital problems, antineutrino energy estimation, the SOI classification, and the IBD classification problem. We begin by discussing the general GA method as it is applied to this problem, and then we examine our feature selections and models. 

\subsection{Genetic Algorithm}

\begin{figure}[!htbp]\centering
\includegraphics[width=1\linewidth]{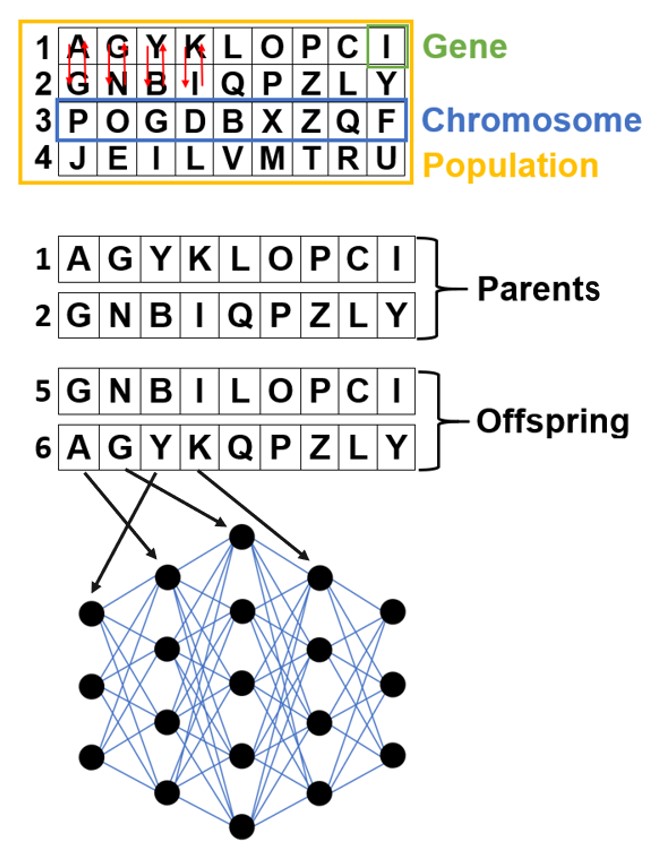}
\caption{Conceptual diagram showing the genesis of neural network construction based on a series of genetic blueprints. The genes will evolve based on a combination of competition and chance.}
\label{f2}
\end{figure}

The heart of the analysis chain begins with a custom genetic algorithm, which holds the genetic blueprints for a variety of deep learning models. These models will compete with each other in a 'survival of the fittest' competition. 

Figure \ref{f2} illustrates the general concept. We begin with a randomized population of 'genes' and each gene carries its own selection of chromosomes which hold specific instructions on how to build particular parts of a neural net. These instructions not only govern the selection of activation functions, learning rates, kernel constraints, the type of feature scaling employed, etc, but also form the neural net structure itself, such as the number of layers and neurons within each layer. For a full list of the available parameters, see Appendix \ref{app}. Moreover, the process is completely automated, as the genes also contain instructions regarding the selection of features themselves (more details are provided in the next subsection). The user first selects a fitness function to determine the quality of a solution, the nature of which depends on whether the target is a classification or a regression. Scores are then assigned to each chromosome, and once a complete generation has had a chance to compete, the top performing genes are selected to mate and create new 'offspring' through the crossing method~\cite{crossover}. We set the probabilities for gene selection for each chromosome in the new population as follows: 40\% chance of a particular gene being selected from the top performer, 25\% from the second best, 20\% from the third best, and 15\% from the fourth best.  We found that the method of 'uniform' crossing ~\cite{crossover} converged to a quality solution the fastest. To introduce a variety in our populations, we also implement a chance mutation for each and every gene before the next generation is due to compete. In this step, the gene is randomly swapped with one of the relevant genes available in the pool. A mutation rate that is too low offers no variety and makes for a slow process, while a mutation rate that is too high creates a random walk scenario. We surveyed mutation rates from 5 to 10 percent, and we found that a mutation rate of 7 percent was optimal for efficient solution convergence.

For regressions, we use a fitness function composed of the R squared ($R^{2}$) score~\cite{https://doi.org/10.1002/bimj.19620040313} minus the standard error of the regression~\cite{stats}, and for the SOI classification we use the accuracy (TP + TN)/(TP + TN + FP + TN) where TP, TN, FP, and FN stands for true positives, true negatives, false positives, and false negatives respectively. The IBD classifier was a particularly challenging problem in part because the approach we took limited the amount of data we could utilize for training (further discussion in Section \ref{IBD Classifier}) and thus we tested a series of fitness functions:   F-score~\cite{https://doi.org/10.2307/1932409}, ~\cite{sorenson1948method} using a $beta$ value of .5, .75, 1.0, 1.25, 1.5. This function aids in ensuring a healthy balance of TP (True Positives) and TN (True Negatives) by penalizing the precision and recall ~\cite{https://doi.org/10.2307/1932409} to various degrees (equally for a $beta$ value of 1). We also attempted the balanced accuracy as well as the precision metric as a fitness function.  We found that the F1 (beta=1)fitness function for the IBD classifier led to models that achieved the best performance on the validation set.

\begin{figure}[!htbp]\centering
\includegraphics[width=1\linewidth]{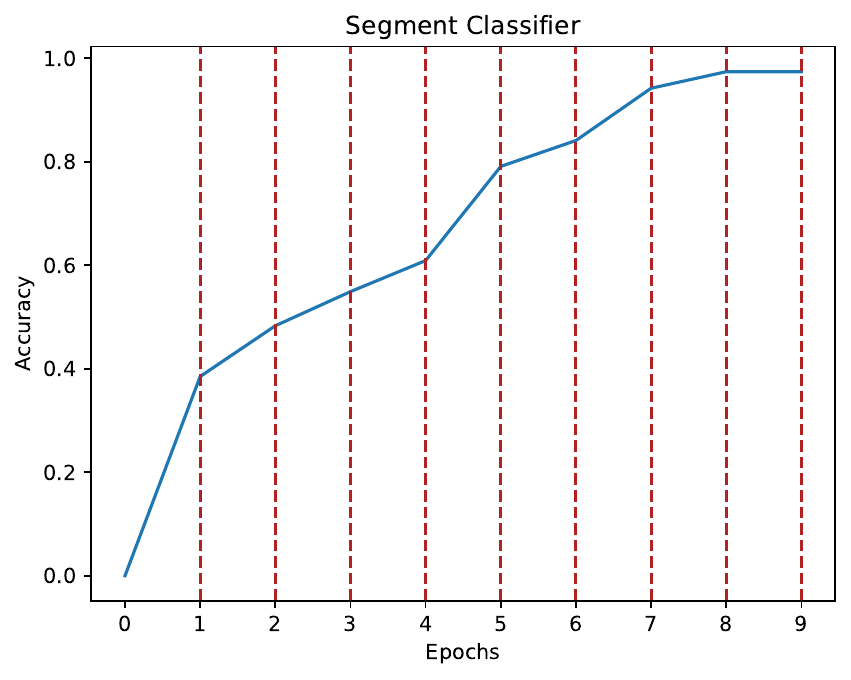}
\caption{Plot showing the rate of solution convergence for a classifier using the GAPE method. Here the classification task is to identify the correct segment where an IBD interaction occurred (referred to in this paper as the SOI classifier). The plot is associated with nine generational epochs and took a total runtime of approximately 33 hours where a population size of 1000 chromosomal arrangements competed.}
\label{f3}
\end{figure}

An example of the GAPE method that creates a classifier from 'genesis' is shown in Figure \ref{f3}. The model attempts to solve the SOI for each IBD event using simulated data for PROSPECT. Here, we started with a very small set of chromosomal arrangements (1000) and ran for eight generational epochs (defined as a
 full cycle through an entire population of chromosomes, not to be confused with training epochs). Further details on the GAPE strategy are provided in the following subsection.  

\subsection{Feature Selection}

Defining a relationship between the total deposited antineutrino energy and the reconstructed energy clusters in the detector is the crux of the higher level analysis done by PROSPECT collaboration's analysis tool (referred to as P2X). A variety of issues make this a difficult problem, including that some of the events are not fully contained in the active materials, the significant amount of passive material in the detector (including non-functional or partially-functional segments), and the time-dependent nature of some detector response parameters, such as scintillation light collection. PG4, a Geant4 based \cite{AGOSTINELLI2003250}  Monte Carlo (MC) simulation of the detector, was developed by the PROSPECT collaboration to model the detector's energy and timing response. A variety of radioactive calibration sources and backgrounds were used to periodically calibrate the detector response.  
IBD MC events are initially generated within PG4 and are then further processed through the P2X analysis chain.  
IBD events were generated assuming a background-free Huber ${}^{235}_{}U^{}_{}$ model \cite{PhysRevC.84.024617}.  
Only events that survive the initial P2X cuts that use the same selection criteria and calibration configuration that was used in \cite{2021} are included for use in this study.  

A list of some of the important features that are part of the output in a typical low level event reconstruction are: the reconstructed longitudinal position ($z$) within each activated segment, the segment hit time defined as $t=\frac{1}{2}(t_0+t_1)$ where $t_0$ is the time of the earliest photon hit in the segment and $t_1$ is the time of the second hit, timing difference between first photon arrivals in the PMTs on each end of a segment $dt=t_1-t_0$, total photon count by each PMT left and right for each segment ($PE_L$, $PE_R$), the total visible energy deposited in each segment $E_{vis}$, ,the segment with the highest energy deposit in an event $E_{max}$, and the PSD in each segment, which is extracted from the waveform shape. These eight features were selected as potential inputs for the ML models created by GAPE. For a full list of P2X outputs, including those not considered for inclusion as ML inputs, see \cite{2019}. 

GAPE has a built-in dimensionality reduction mechanism, in which less useful features are systematically weeded out. The reduction is generated by forward feeding a block of features into an adaptive deep learning model, where a fitness function evaluates the precision of the model. Features that produce a more precise model are more likely to be inherited by the next generation of candidates. 

The process of finding a solution is divided into two parts. In the first stage, we use the GA not only to evolve the entire parameter set but also to select the right features through dimensionality reduction. Too many features can make a model overly complex and slow down the learning process and too few features can lead to underfitting. We use a smaller subset of our selected simulated IBD MC datasets for the SOI and energy estimator in the GA process to avoid unpredictable memory overload failures (200000 events for the training set and 40000 events for validation).  

Data for the SOI and energy estimators were collected as a 154 (defined by the total number of PROSPECT segments) by X-features array, which was flattened before being forward fed at the inputs. Each row represented a specific channel, while the X columns represent the X select features. For the IBD classifier, we collect a 154 by X$*$2 array with the first 154 by X block containing the prompt event information, and the second 154 by X block containing the delayed event information. The array is then flattened prior to being forward fed into the network. The hit time $t$, the left and right photon count, ($PE_L$, $PE_R$), and the PSD are fixed and never removed.  The portion of the chromosome that governs feature selection contains 5 genes that generate a random integer from [0,1]. A value of 0 instructs the GAPE to zero out the column associated with that gene and a value of one instructs GAPE to keep the column unchanged. The features are then scaled according to the instruction dictated by the gene associated with scaling. For stage 1 we choose 1000 random chromosomes and run each of the 1000 candidate models for 2 epochs. The models compete for a total of 4 generational epochs. After the fourth and final generational epoch, the best performing model of all is selected along with the features associated with that model. In stage 2, the genes for selecting features are turned off, and all future candidate models inherit the fixed features from the winner from stage 1.  We choose 500 candidate models for the first generational epoch of the new stage, and 200 for the remaining generational epochs. We also include the winning model's architecture into a fresh population set but we do not mate it with the other candidates at the start of stage 2. 

In stage 2, each candidate model is trained three times per generational epoch with randomized batch shuffling, totaling six training epochs, and the average score across these trials is computed. The same was done for the IBD classifier (except here we use a smaller specialized dataset explained in Section~\ref{IBD Classifier}) and the average value from a k-fold crossing~\cite{article_k} with 5 folds was used for scoring. The classifier with the highest score is then selected. We recommend k-fold crossing averaging as the standard for future runs as it provides the most comprehensive survey of a model's performance albeit with a trade-off of longer runtimes. The total number of generational epochs in stage 2 is determined when it is found that the metrics of the top model does not improve upon the previous generation's winner. For all three analysis tasks in stage 2, we found that the best solution was achieved by the 4th generational epoch. Once the optimal architecture of each model is determined by GAPE, the final models for the energy estimator and the SOI classifier are evaluated using a k-fold crossing with a total of 26 and 30 epochs respectively using the full dataset discussed in Section \ref{Machine Learning}.

 For each model considered in this paper, features $t$, $PE_L$, $PE_R$, and PSD were eventually selected by the GAPE in each trial. These quantities are foundational for all other estimated values. The PSD helps to identify particle type, and in conjunction with $PE_L$ and $PE_R$, we can reconstruct the deposited energy in each segment along with the point of interaction via relative amplitude information. Along with $t$, which allows us to order interactions in time within each event, the machine learning algorithm has the necessary features to learn how to reconstruct information such as the incoming antineutrino energy.

\subsection{Machine Learning} \label{Machine Learning}

We use a multi-layer function-fitting feed-forward network constructed via the GAPE method, as seen in Figure \ref{f1}.  We construct our models using both PyTorch 0.4.0~\cite{pytorch} and Tensorflow 2.6.0~\cite{tensorflow2015-whitepaper} with equivalent performance. The full parameter space explored by GAPE is provided in the Appendix \ref{app}.

For the energy estimator, the GAPE settled on a 6 layer feed-forward neural net: A 616-length input layer containing the four selected features for each segment is connected to 4 hidden layers of size 900, 700, 350, and 250, and one linear output. The respective activation functions selected by the GAPE on the hidden layers are the rectified linear unit function (relu) \cite{inproceedings}, exponential  linear unit function (elu) \cite{clevert2016fast}, relu function, and scaled exponential linear unit (selu) \cite{klambauer2017selfnormalizing} function respectively. The ADAGRAD optimizer~\cite{article} was selected with a learning rate of .01, initial accumulator value of 0.1, and an epsilon of $10^{-7}$. Global initialization of hidden layer weights was done using a Glorot Uniform initializer~\cite{pmlr-v9-glorot10a}. A mean squared error loss function and a batch size of 16 were selected. Finally, the GAPE method selected a solution that involved no scaling of features.  

The SOI classifier required a less complicated structure. The GAPE method settled on a 4-layer feed-forward neural net: A 616-length input layer containing the four selected features for each segment is connected to 2 hidden layers of size 475 and 450, and a softmax output of size 154. The hidden layer activations selected were the rectified relu function, and the hyperbolic tangent function (tanh) \cite{steeb2005nonlinear} respectively. A learning rate of .0001, a categorical cross-entropy loss function~\cite{cybenko1998mathematics}, the Adam optimizer~\cite{kingma2017adam} (exponential decay rates for the first and second momentum of 0.9 and 0.999 respectively, $\epsilon= 10^{-6}$), and a batch size of 64 were selected. Glorot Uniform initialization of the hidden layer weights was selected. Finally, the GAPE selected to scale the features using the standard scaler~\cite{scikit-learn}.  

For an initial IBD classifier which we refer to as `Classifier 1,' the balanced accuracy fitness functions performed the best overall. GAPE selected a 6-layer feed-forward neural net: A 616-length input layer containing the four selected features for each segment is connected to 4 hidden layers of sizes 1120, 160, 240, and 320, and a sigmoid output. The hidden layer activations selected were the rectified relu, elu, selu and tanh functions respectively. A learning rate of 0.0001, a binary cross-entropy loss function, the Nadam optimizer \cite{Dozat2016IncorporatingNM} (exponential decay rates for the first and second momentum of 0.9 and 0.999 respectively, $\epsilon= 10^{-7}$), and a batch size of 40 were selected. Kernel initializers selected for the hidden layers were Lecun uniform \cite{article_lecun}, truncated normal \cite{tensorflow2015-whitepaper}, glorot normal \cite{tensorflow2015-whitepaper}, and he normal \cite{he2015delving} respectively. Finally, the GA selected to scale the features with the standard scaler.

Another classifier was trained on a smaller subset of data (the purpose of which is described in the results section) which we refer to as `Classifier 2.' The F1 fitness function performed the best overall (balanced accuracy was competitive). Here, the GAPE algorithm selected a  6-layer feed-forward neural net: A 616-length input layer containing the four selected features for each segment is connected to 4 hidden layers of sizes 80, 480, 160, and 240, and a sigmoid output. The hidden layer activations selected were the relu, tanh, elu, and elu functions respectively. A learning rate of .0001, a binary cross-entropy loss function, the Nadam optimizer (exponential decay rates for the first and second momentum of 0.9 and 0.999 respectively, $\epsilon= 10^{-7}$), and a batch size of 100 were selected. Kernel initializers selected for the hidden layers were orthogonal \cite{tensorflow2015-whitepaper}, truncated normal, truncated normal, and he uniform \cite{he2015delving} respectively. Finally, the GA selected to scale the features with the standard scaler.

For GAPE we used a smaller subset of 240000 events which we divided in the following way:

\begin{enumerate}
\item {\bf Train (80\%).} Used during training to update network parameters (weights and biases).
\item {\bf Validate (20\%).} Used to stop training to prevent overtraining and is not used to update network parameters.

\end{enumerate}

Once GAPE selects the best model architecture, we train the very final models, used for evaluation, on a larger dataset comprised of approximately 5.5 million
simulated IBD candidates. This larger dataset was split into three groups:

\begin{enumerate}
\item {\bf Train (70\%).} Used during training to update network parameters (weights and biases).
\item {\bf Validate (10\%).} Used to stop training to prevent overtraining and is not used to update network parameters.
\item {\bf Test (20\%).} Used to evaluate performance and left completely separate. This set is not used during training at all.
\end{enumerate}

The dataset for the IBD classifiers was handled differently and will be described in \ref{IBD Classifier}.

Supervised learning was performed on the Lawrence Livermore National Laboratory  Computer Cluster (Lassen) and on a Google Cloud Platform (GCP) virtual machine utilizing a Nvidia Tesla V100 GPU.

\section{Energy and Position Estimator Results and Performance} \label{Sim Perf}

In this section, we explore the GAPE-selected machine learning (ML) models performance using simulated IBD MC events and compare them with previous models when possible. All events analyzed in this section are considered IBD candidates as selected by P2X using the same selection criteria and calibration configuration used in \cite{2021}. 

The positron from an IBD interaction in the detector typically deposits up to 8 MeV across a number of segments (1-3 usually). For the prompt event, IBD selection requires a total reconstructed visible energy between 0.8 and 7.2 MeV and individual reconstructed pulse PSD values that are all all within $\pm$2$\sigma$ of the PSD mean of the calibrated electron-like particles. As the IBD neutron thermalizes it produces negligible scintillation, and within approximately a 50 $\mu$s time constant and 15cm of the IBD vertex point it captures.  Any further cuts to the candidate selection for any of the particular models will be detailed in each subsection. When a neutron captures on ${}^{6}_{}Li^{}_{}$ (occurs approximately 75 percent of the time) a ${}^{3}_{}H^{}_{}$-${}^{4}_{}He^{}_{}$ pair will produce 0.526 MeV of total visible energy that is emitted within a few micrometers of the capture vertex. For selection, the energy must be within $\pm$3$\sigma$ of the mean value and the PSD must be within $\pm$2$\sigma$ of the mean value (determined using cosmogenic n-Li capture events).  This delayed event must occur within 120 $\mu$s and no sooner than 0.8$\mu$s of the prompt cluster for selection. The capture segment must be the same as that of the prompt cluster or vertically/horizontally adjacent to it. For a more detailed survey of the P2X analysis chain and selection criteria see \cite{2021}.

\subsection{SOI Classifier}

The SOI classifier attempts to identify and label events that occur in dead segments as well as live segments. The GAPE selected model was frozen after training for 16 epochs and inference on the test set produced the results in this section. We found improvements in accuracy under 3 different scenarios. When only the initial P2X analysis cuts are applied, we refer to this as the \textit{no additional cuts} scenario. All events are subject to the P2X initial analysis cuts. We define the additional removal of events in which IBD interactions occur in dead material or dead segments as \textit{DS cuts}. \textit{Low energy cuts} are defined as the removal of events in which the total visible energy in the detector is less than 0.8 MeV.  For the no additional cuts, low energy cuts, and the low energy plus DS cuts scenarios, respectively we obtained (95.4, 98.0, 98.7)$\%$ accuracies for the ML SOI classifier and (92.4, 97.2, 97.9)$\%$ accuracies for the traditional method to determine the SOI. A k-fold crossing analysis was done using the entire train and validation sets and k=10 and the results are tabulated in table \ref{table:1}.  

\begin{figure}
\includegraphics[width=1\linewidth, height=13cm]{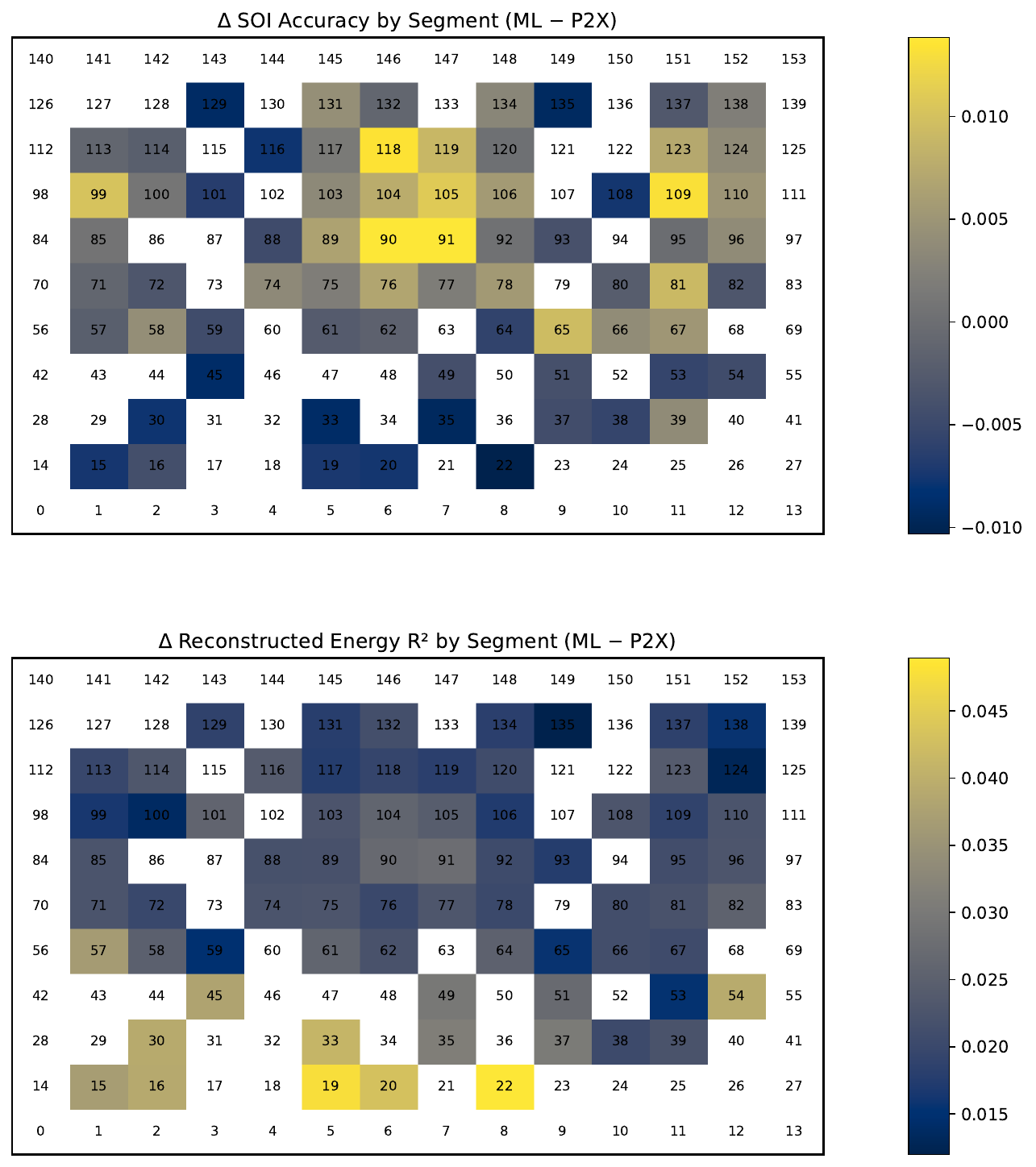}
\caption{Shown is the relative performance of the ML and traditional models on the test set by segment. Integer values in the grid represent segment numbers. While the overall performance is better with the ML model, we see in the top plot that the SOI ML classifier does best in segments with more active neighbors, while the P2X method does better in the more isolated segments. For the reconstructed energy, in the bottom plot we observe similar performance gains across all segments with the greatest gains seen in the more isolated segments.}
\label{f4}
\end{figure}

The ML model yields modest gains in SOI reconstruction relative to an already well-performing traditional reconstruction technique when low energy cuts are applied (these cuts were implemented in the final analysis in \cite{2021}). However, it is noteable that the gains of the ML model are significant when the low energy cuts are not applied. The relative performance of the SOI classifier by segment when low energy cuts are applied is shown in the top of Figure \ref{f4}. The color maps represent the average accuracy for each segment, calculated by subtracting the P2X method from the ML method. We see that the biggest gains in performance for the ML models are in segments surrounded by active neighbors, while the more isolated segments perform better using standard reconstruction techniques.

\begin{table} [!htbp]
\fontsize{10}{10} \selectfont \centering
\begin{tabular}{l|lllll}
method		& No Cuts 	& Low Energy Cuts 		& Low Energy \\ & & & and DS cuts 			\\
        
\hline 	
 P2X	& 92.4 $\pm$ 0.5			& 97.2 $\pm$ 0.3  			&  97.9 $\pm$ 0.2 				\\

\bf ML	& \bf 95.4 $\pm$ 0.3		& \bf 98.0	$\pm$ 0.2	    & \bf 98.7	$\pm$ 0.1	\\
\end{tabular}
\caption{A MC study measuring accuracy. We see the deep learning model performs better than the traditional model for SOI reconstruction. This table summarizes improvements as they relate to no cuts ( beyond primary P2X candidate cuts), low energy cuts, and both low energy and DS cuts. }
\label{table:1}
\end{table}

\subsection{Energy Estimation} \label{energy}

\begin{figure}
\includegraphics[width=1\linewidth, height=13cm]{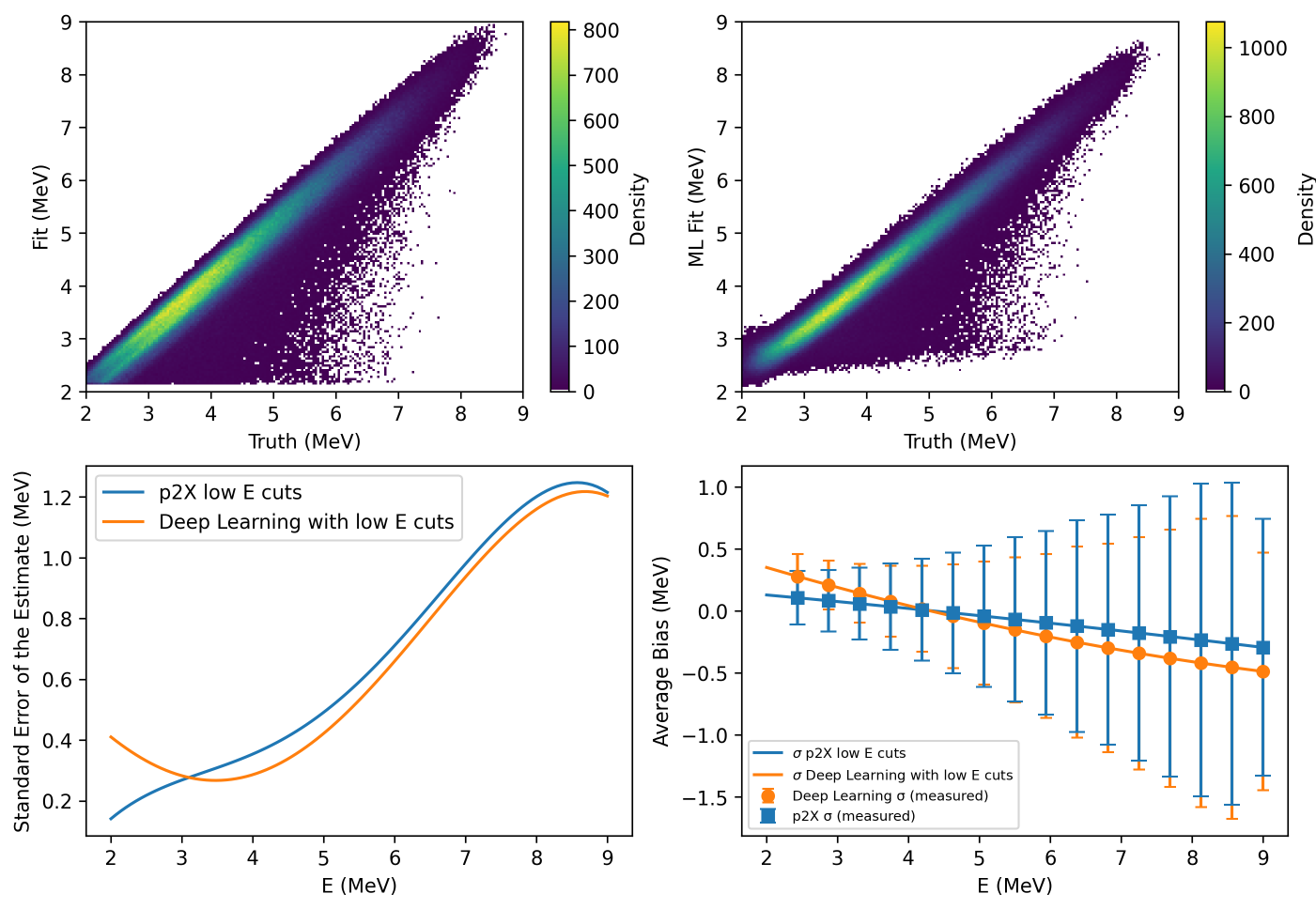}
\caption{Performance comparison for antineutrino energy estimation. Results are displayed for traditional model (top-left) and the ML energy estimator (top-right). Low E cuts are applied. We see the fitted values match the underlying truth better overall using the ML model. The bottom left plot shows the standard error of the estimate where we see that the the predicted values in the ML model differ less from the actual values except for energies less than 3 MeV. The bottom right plot shows the average bias (reconstructed minus truth)  and the standard deviation across the energy spectrum where we generally see a tighter bias range for the ML model.}
\label{f5}
\end{figure}

Figure \ref{f5} shows the reconstructed antineutrino energy versus the truth as reported in PG4. Model performance is evaluated using the $R^{2}$ score, which measures the fraction of variance in the dependent variable explained by the true energy. Low energy cuts are applied in both analyses. In the top-left plot, we minimized the $R^{2}$ score by adding a best-fit constant to all the prompt energy values.  
It is important to note that P2X makes no effort to attempt any corrections for missing energy - those corrections are done statistically using an unfolding process and are beyond the scope of this paper.  
Additionally, while the ML energy estimator does the unique workload of predicting the true antineutrino energy, no further efforts are made to statistically unfold the distribution the model yields. 
Thus one-to-one comparisons are somewhat grainy, but by shifting the reconstructed P2X-outputted reconstructed visible energy values to best match the true energy, we can attempt to quantify and compare the unexplained variance of each representation.  

For this particular test set, the traditional model had an overall $R^{2}$ score of 0.856. (0.874 with DS cuts). The top-right plot shows the performance of the GAPE-selected ML model (described in the previous section) in estimating the true energy. The same test set was used for both plots. The ML model outputs an overall $R^{2}$ score of 0.876 (0.891 with DS cuts). We qualitatively see that the ML model proves to be an overall better estimator of the true energy than the traditional model by how tightly the estimator follows the true value. To make this comparison more explicit, the bottom left plot compares the standard error of the estimate for each method. In this comparison we see that the ML method provides superior reconstructed energy resolution across the energy spectrum above approximately 3 MeV. Both plots exhibit peak performance at approximately 3.5 MeV. The bottom right plot displays the bias (reconstructed minus truth) in reconstructed energy, with the error bars representing the standard deviation. We find that both bias and resolution metrics indicate peak performance of the ML algorithm at approximately 3.5 MeV true neutrino energy.

The relative performance of the energy estimator performance by segment when low energy cuts are applied is shown in the bottom of Figure \ref{f4}. The color maps represent the average $R^{2}$ for each segment, calculated by subtracting the P2X method from the ML method. Contrary to the SOI estimator, performance gains are observed across all segments, with the greatest gains seen in the more isolated active segments.

 To evaluate the model's ability to generalize, a k-fold crossing analysis (with k=10) was done using the entire train and validation sets and the results are tabulated in table \ref{table:2}. The greatest performance for the ML energy estimator gives an $R^{2}$ score of 0.892 $\pm$ .001 while the greatest performance gap is provided in the conditional absence of additional cuts.

 \begin{table} [!htbp]
\fontsize{10}{10} \selectfont \centering
\begin{tabular}{l|lllll}
method		& No Cuts 	& Low Energy Cuts 		& Low Energy \\ & & & and ML cuts 			\\
        
\hline 	
 P2X	& 0.63 $\pm$ .04			& 0.853 $\pm$ 0.003  			&  0.873 $\pm$ 0.002 				\\

\bf ML	& \bf 0.81 $\pm$ .3		& \bf 0.877	$\pm$ .002	    & \bf 0.892	$\pm$ 0.001	\\
\end{tabular}
\caption{A MC study comparing $R^{2}$ scores. We see the deep learning model performs better than the traditional model for energy estimation. This table summarizes improvements as they relate to no cuts ( beyond primary P2X candidate cuts), low energy cuts, and both low energy and DS cuts. }
\label{table:2}
\end{table}

\section{IBD Classifier} \label{IBD Classifier}

\subsection{Introduction}

The GAPE method was used to select a classifier for the identification of IBD interactions from backgrounds, in contrast to a typical background subtraction. For an input PROSPECT prompt-delayed event pair, the classifier outputs a single score on a continuous scale between 1 (more IBD-like) and 0 (less IBD-like).  To make a final selection, event pairs with a score greater than 0.5 are classified as signal events, while events with scores below that are classified as background. This classifier is meant to be implemented on event pairs that have successfully passed the traditional selection as applied through the P2X analysis chain. We take a unique approach here by using real backgrounds for training. Because the decision process is opaque in ML models, it becomes challenging to identify or mitigate errors and biases that may be present. These errors can be introduced in the training process itself through errors in background modeling. The motivation for our unique approach is to remove these errors in the training data and thereby help mitigate this aspect of the black-box problem.

In order to approximate the proportions of backgrounds and the estimated number of antineutrinos in our training and validation sets, we carefully select a mixture of real-world reactor-off correlated cosmic backgrounds and reactor-on accidental backgrounds in such a way that the proportions match the measured final IBD candidate set composition in \cite{2021}. The reactor-off correlated cosmic backgrounds are events that pass the initial P2X candidate selection with the additional removal of all events where the total deposited energy is less than 0.80 MeV or greater than 7.2 MeV. For the reactor-on accidentals, we include the same P2X selection cuts and low and high energy cuts but with the reverse exception to include only the events that fall outside of the correlated timing window.  These accidental background events and reactor-off correlated cosmogenic backgrounds are described in detail in \cite{2021}. 

In the following sub-section, we detail the steps taken to develop a first IBD classifier, which we term `Classifier 1.' A follow-up IBD signal classifier, referred to as 'Classifier 2,' is discussed afterwards.  For Classifier 1, we randomly select 21166 reactor-on accidental events from our sample and perform a data augmentation by artificially resetting the delayed signal arrival times with random samples from a uniform distribution ranging from 1 and 120 microseconds (the IBD signal selection’s required time coincidence window). We use the full set of reactor-off correlated cosmogenic backgrounds (30568 events). These samples were combined with 40054 antineutrino events simulated with the Huber spectrum to maintain the expected balance of antineutrinos with backgrounds in the final set of IBD candidates selected by P2X. 

One of the limitations encountered for this method in this experiment is that, while we have virtually unlimited amounts of simulated antineutrino data and a plethora of reactor-on/off accidentals, we have an underwhelming amount of available reactor-off correlated cosmogenic events for training/testing. We refer to this uniquely treated mix of simulated neutrinos and real backgrounds, meant to mimic the mixture of candidates thought to represent the final IBD candidates in this experiment, as the 'mixed bag'. Given the undersized dataset at hand, we elected not to reserve an isolated test set. As such, for the development of Classifier 1 the mixed bag was divided as 80 percent training, and 20 percent validation. 

\subsection{Initial Results}

When Classifier 1 is applied to the mixed bag validation dataset, this dataset's true IBD signal and true IBD background components are classified as signal-like (a `positive' designation) and background-like (a `negative' designation) categories.   
The signal-like classified samples consist of true simulated IBD events (TP) and true backgrounds (FP); false negatives (FN) and true negatives (TN) are likewise sub-components within the background-like classified samples.  
Applying k-fold crossing (k=5) with Classifier 1 on the different folds of the validation set produces an average recall, (more colloquially referred in particle physics as signal selection efficiency), TP/(TP+FN) of 78.5 $\pm$ 3\%. This 21.5\% reduction in IBD signal statistics is smaller than the 60-70\% reduction introduced in the preceding conventional IBD pre-selection cut stage.  We define the Predictive Positive Rate (PPR), i.e., the selected fraction, as $\frac{TP+FP}{TP+FP+TN+FN}$ and find an average of 47 $\pm$ 3\%.  Meanwhile, for the same k-fold crossing, the average precision, $\frac{TP}{TP+FP}$,  is 73 $\pm$ 1\%, and the ratio of true IBD events to backgrounds in the signal-like classified sample, TP/FP (also referred to as signal-to-background ratio, or SBR), is 2.76$\pm$0.23.  
This is a substantial improvement ($\approx 3.6\times$) over the 0.77 ratio achieved by the IBD pre-selection cuts.  Both of these outcomes are promising first indications of IBD classifier performance. 
Additionally, we see an improvement of 32 $\pm$ 2$\%$ in the effective statistics \cite{Cowan_2011},  defined as $\frac{S^2}{S+B}$ where S and B are the total number of TPs and FPs in the set, over the original IBD pre-selection set.  
Performance results are also summarized in Table~\ref{table:3}.  

\begin{table}[ht]
\centering

\begin{tabular}{|p{6.0cm}|p{2.5cm}|p{2.5cm}|}
  \hline
  \textbf{Metric} & \textbf{Classifier 1} & \textbf{Classifier 2} \\
\hline
  \hline
  Precision & 73 $\pm$ 1\% & 70.5 $\pm$ 0.8\% \\
  Recall & 78.5 $\pm$ 3\% & 71 $\pm$ 3\% \\
  F1 & 75.8 $\pm$ 0.5\% & 70 $\pm$ 1\% \\
  Accuracy & 78.1 $\pm$ 0.4\% & 72.5 $\pm$ 0.7\% \\
  Balanced Accuracy & 78.1 $\pm$ 0.3\% & 72.4 $\pm$ 0.8\% \\
  Signal-to-Background & 2.8 $\pm$ 0.2 & 2.4 $\pm$ 0.1 \\
  Effective Statistics Improvement & 32 $\pm$ 2\% & 7 $\pm$ 4\% \\
  Selected Fraction (Mixed Bag) & 47 $\pm$ 3\% & 46 $\pm$ 3\% \\
    \hline
  Selected Fraction (Data) & 31 $\pm$ 2\% & 40 $\pm$ 2\% \\
    \hline
\end{tabular}
\caption{Performance metrics for IBD Classifiers 1 and 2.  Metrics for most rows refer to performance of the classifier on its relevant mixed bag validation sample, while for the last row performance on the relevant PROSPECT dataset is specified.}
\label{table:3}
\end{table}

\begin{figure}
\includegraphics[width=1\linewidth]{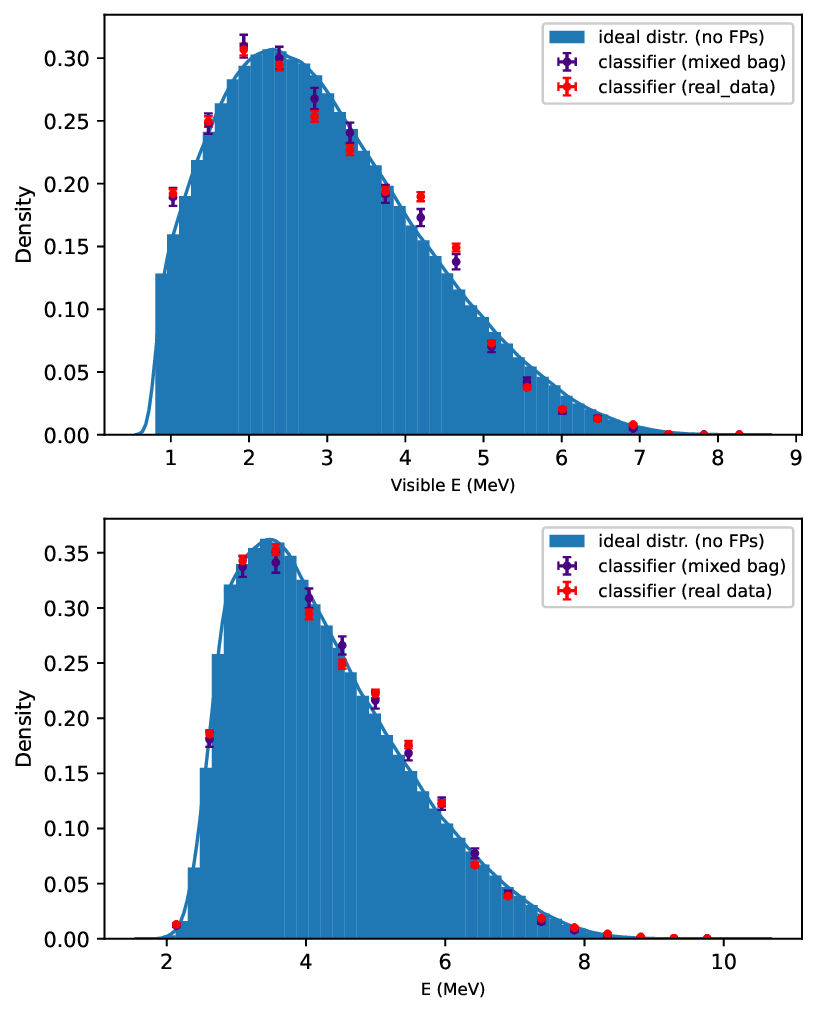}
\caption{ Plots of the mixed bag ML IBD signal distributions and the real-word ML IBD signature distributions in both the prompt energy space (leftside) and the ML energy space (rightside).} 
\label{f6}
\end{figure}

Figure~\ref{f6} depicts the energy spectrum of the signal-like classified sample within the mixed bag using either the conventional IBD prompt energy estimator or the ML-derived neutrino energy estimator presented in Section~\ref{energy}.  
We again emphasize that the comparison between energy metrics is not strictly one-to-one: the ML model predicts true antineutrino energy, while the traditional method reconstructs visible energy without unfolding corrections.  
The comparison is therefore intended as a qualitative indicator of performance rather than a direct equivalence.  
In either of these spaces, we see the energy distribution of the signal-like classified sample as the indigo data points. The solid blue bars represent the expected energy spectrum of a pure sample of true IBD signal events, which is generated by plotting the energy value of true IBD events from a large IBD MC simulation sample using the same underlying Huber ${}^{235}_{}U^{}_{}$ reactor neutrino model \cite{PhysRevC.84.024617} used to generate the mixed bag's true IBD component.  
All distributions in Figure~\ref{f6} are manually set to have equivalent normalizations.  
In the unrealistic but ideal case where our classifier contains no FPs we would expect the indigo data points in each plot to overlap their respective Huber-derived distribution in filled blue.  
In practice, the mixed bag signal-like classified sample contains expected deviations from the pure-TP spectrum due to the presence of residual backgrounds.  
In particular, the mixed bag's signal-like energy spectrum is higher than that of a pure IBD sample
in the 4-5 MeV prompt energy regime due to the presence of a peak in the correlated background sample arising from fast neutron inelastic scatters on $^{12}$C nuclei~\cite{2021}.  
In addition, the mixed bag's signal-like energy spectrum (which we refer to as the ML IDB signature distribution) contains an excess at low prompt energies, since residual accidental backgrounds are expected to inhabit this energy range.  As expected, similar features are also seen on when comparing in ML energy space.

Beyond demonstrating the efficacy of Classifier 1 in improving PROSPECT's IBD selection, we must also probe classifier performance on real PROSPECT datasets containing a mixture of IBD signals and backgrounds.  
The value of this demonstration for the IBD classifier is distinct from a potential data-based demonstration for the SOI and energy estimators described in Section~\ref{Sim Perf}.  
In these cases, performance of the ML estimator can be compared directly to its conventional estimator counterpart, whose performance has already been validated on real data in past PROSPECT publications.  
Since the IBD classifier is designed to be applied to the output of the conventional IBD selection, such a direct performance comparison is not possible.  

The presence of bias in the IBD classifier can be easily conveyed by comparing the selected fraction of events that are classified as signal-like in the mixed bag validation sample and in the full PROSPECT reactor-on dataset.  
From the same k-fold crossing described above, average fractions are 47\% and 31\%, respectively, indicating that a substantial bias is indeed present in Classifier 1.  
In comparing the normalized energy spectra between the mixed bag and true PROSPECT reactor-on datasets in Figure~\ref{f6}, however, one does not see a large energy-dependent deviation between the two datasets.  
This suggests a bias origin that is not primarily related to the IBD positron signal that forms the basis of the neutrino energy estimate.

A variety of factors could potentially explain the observed bias in signal-like classification between validation datasets and real reactor-on datasets.  
Some possibilities, such as the presence of previously-unaccounted-for correlated reactor-on backgrounds, seem unlikely based on prior PROSPECT studies~\cite{2021}.  
Another potential cause could be the training of a rigorous classifier with an emphasis on being precise, albeit selective: this would result in a model that could reject any edge-case events that diverged from expectations. This effect would result in a higher FN count than what is observed in the mixed bag.  
A plausible final possibility stems from the use of MC simulated IBD events in place of real IBD events in the mixed bag training dataset.  
A lack of fine details in the simulation (e.g. non-precise neutron transport and capture inputs, incomplete simulation of scintillation photon production and transport or electronics response, use of calibration constants outside the range of validity) or inaccurate descriptions of these details in the simulation, could cause simulated IBD events to differ meaningfully in their appearance from true IBD events.  
This could result in a training process that emphasizes and attempts to exploit signal-background differences that are not actually present in real PROSPECT data.  

\subsection{Classifier Adjustments and Follow-Up Investigations}

One well-established attribute of real PROSPECT data that is not included by default in PROSPECT's MC simulations is the presence of varying response throughout the detector's lifetime.  
In particular, it was observed the total light collection and effective attenuation length of its active detector volume slowly decreased over the course of PROPSECT running, with the former decreasing to roughly 60\% of its initial value from the start to end of data-taking~\cite{2021}.  
Time-dependent changes in light production or propagation had downstream effects on other important physics features, such as PSD parameter central values and band widths, segment hit time offset $dt$ distributions, and reconstructed longitudinal positions $z$.  
Various strategies were implemented to calibrate time-dependent response attributes, such that primary physics features or IBD selection cut values could be defined in a time-independent manner.  
For example, time-dependent absolute scaling and Gaussian smearing factors determined by the location and width of the n-Li capture peak were applied to PROSPECT per-segment E$_{vis}$ features to ensure consistency in energy reconstruction for all PROSPECT data~\cite{PROSPECT:2018dtt}.  
On the other hand, some downstream attributes, such as PSD features and $z$ widths, were not similarly time-equalized, and instead modest time-dependencies were treated with a time-dependent conventional IBD selection cut scheme.  
These differences in calibration and feature definitions open the door for time-dependent residual feature differences between data and MC IBD events that could be responsible for the training biases observed in Classifier 1.  

To investigate this potential source of bias, we selected a smaller subset of reactor-on/off data to train an IBD classifier model termed Classifier 2.  
We select data from 'period 2' as defined in \cite{adriamirado2023final}, which constitutes a time interval of 22.8 (15.7) days of reactor-on (-off) data recorded during the first half of PROSPECT data-taking~\cite{PROSPECT:2022wlf}.  
Since detector response variations in PROSPECT occurred gradually over time, the use of a more abbreviated data period for classifier training should reduce biases related to time-dependence.  
This period was selected because PROSPECT's default simulated detector response utilizes calibration data relevant to the calendar range that defined period 2.  
This period-specific approach to training should not to be confused with the recent PROSPECT spectral and sterile analyses based on a multi-period, configuration-aware framework~\cite{adriamirado2023final}, which leverages new single-ended segment reconstructed data features and PROSPECT's conventional selection criteria, which is explicitly designed to preserve response and IBD spectrum shape fidelity across all detector states and periods.  

The P2X analysis chain and conventional IBD selection for this period yields a total of 27163 events. The same analysis of the reactor-off set yields a total of 6519 correlated cosmogenic backgrounds.  
After live-time scaling, we estimate a mixture distribution of approximately 46$\%$ antineutrinos, 35$\%$ cosmogenic backgrounds, and 19$\%$ reactor-on accidentals from the period 2 reactor-on set.  
To compose the mixed bag sample for Classifier 2, we select 8695 simulated neutrinos, 3473 augmented reactor-on accidentals (identical to the augmentation in the Classifier 1 mixed bag), and a full set of 6519 correlated reactor-off events.  
The signal-to-background ratio (SBR) delivered by the conventional IBD selection for this period is thus $\frac{8695 \pm 70}{3473 \pm 4 + 6519 \pm 30} = 0.870 \pm .008$.  
Errors represent statistical uncertainties in the relevant signal and background datasets.  

Performance metrics for Classifier 2 are listed in table \ref{table:3}.  
Using the same k-fold crossing approach as Classifier 1, Classifier 2 delivers an average precision (IBD detection efficiency) of 70.8$\pm$0.8\%, slightly worse than that delivered by Classifier 1.  
The improvement in effective statistics over the IBD pre-selection set now stands at a lower 7$\pm$4\%.    
Meanwhile, Classifier 2 achieves a SBR of 2.4$\pm$0.1.  
Although this represents less of an improvement with respect to Classifier 1 on its relevant dataset -- Classifier 1 improves this ratio from 0.77 to 2.8 for the full PROSPECT dataset while Classifier 2 improves it from 0.87 to 2.4 for period 2 -- apparent gains in background reduction are still clear and substantial. 
The significant increase in SBR suppresses systematic uncertainties from background modeling and fluctuations, and for SBR $>$ 2, offers the potential to enable testing for signal deviations using a signal-plus-background framework without explicit background subtraction \cite{BIEKOTTER2021136311}.  
Most importantly, the signal-selected fraction for Classifier 2 is 40$\pm$2\% for the mixed bag and 46$\pm$3\% for period 2 PROSPECT data. This represents a major reduction in bias with respect to Classifier 1, indicating that time-dependent offsets in PROSPECT physics features between data and MC IBD events were indeed primarily responsible for that classifier's large observed bias.  
While the data-period-specific training mitigates observed biases, we note that the degree to which the resulting models generalize across detector conditions has not been fully explored and remains an area for future study.

\section{Conclusion}

We have developed and demonstrated a novel approach called GAPE which utilizes a genetic algorithm to perform feature selection for and seed and evolve deep learning solutions to physics metric reconstruction and signal selection in a reactor neutrino dataset.  
This method can be generalized to a wide range of problems in the analysis of nuclear and particle physics experimental datasets and other fields.  
Using GAPE to guide the construction of optimized deep learning models, we obtain an energy and position estimator, as well as an antineutrino classifier for the PROSPECT detector. 
When contrasted with prior efforts reported in the 2021 PROSPECT analysis~\cite{2021}, which employ traditional statistical approaches and use the same underlying dataset, our results show several notable improvements. 
We observe modest improvements in the accuracy of position estimation and modestly higher $R^2$ scores for energy estimation.  
Observed improvements in segment position reconstruction are comparable in scope to prior AI/ML-based studies of position reconstruction along PROSPECT cell lengths~\cite{andriamirado2025machinelearningsingleendedevent}.  

Most notably, the developed IBD classifier offers the promise of substantially improving the signal-to-background ratio for PROSPECT's centerpiece IBD neutrino interaction signature.  
When applied in conjunction with the existing P2X selection, the GAPE-selected models provide a noticeable improvement in background rejection relative to the traditional approach: relative to the conventional IBD analysis considered in this study, Classifier 2 improves this ratio by nearly 2.8 times.  
Meanwhile, it is shown that a data-period-specific IBD classifier training regimen is necessary to avoid large classification biases arising primarily from the time-dependent nature of some model input physics features. The full development and quantitative assessment of performance of an unbiased IBD classifier on real-world reactor neutrino data will require increased reactor-off training dataset volumes, improved period-specific IBD MC training dataset preparation, and a more detailed examination of the classifier’s efficiency and bias as a function of energy.  
Future work could also be directed towards extending the usage of GAPE AI/ML techniques to PROSPECT physics analysis processes downstream from IBD selection, such as the comparisons between measured and predicted neutrino energy spectra.

\acks{``This material is based upon work supported by the following sources: US Department of Energy (DOE) Office of
Science, Office of High Energy Physics under Award No.
DE-SC0016357 and DE-SC0017660 to Yale University, under Award No. DE-SC0017815 to Drexel University, under
Award No. DE-SC0008347 to Illinois Institute of Technology, under Award No. DE-SC0016060 to Temple University,
under Award No. DE-SC0010504 to University of Hawaii,
under Contract No. DE-SC0012704 to Brookhaven National
Laboratory, and under Work Proposal Number SCW1504 to
Lawrence Livermore National Laboratory. This work was performed under the auspices of the U.S. Department of Energy
by Lawrence Livermore National Laboratory under Contract
DE-AC52-07NA27344 and by Oak Ridge National Laboratory under Contract DE-AC05-00OR22725. Additional funding for the experiment was provided by the Heising-Simons
Foundation under Award No. 2016-117 to Yale University.
J.G. is supported through the NSF Graduate Research
Fellowship Program. This work was also supported by
the Canada First Research Excellence Fund (CFREF), and
the Natural Sciences and Engineering Research Council of
Canada (NSERC) Discovery program under grant No. RGPIN418579, and Province of Ontario.
We further acknowledge support from Yale University, the
Illinois Institute of Technology, Temple University, University
of Hawaii, Brookhaven National Laboratory, the Lawrence
Livermore National Laboratory LDRD program, the National
Institute of Standards and Technology, and Oak Ridge National Laboratory. We gratefully acknowledge the support and
hospitality of the High Flux Isotope Reactor and Oak Ridge
National Laboratory, managed by UT-Battelle for the U.S. Department of Energy"~\cite{2021}}

\appendix
\section{Full Parameter Space For Gape}
\label{app}

\addtocounter{section}{-1}
\renewcommand{\thesection}{}

The tables below summarize the genetic blueprints that were available to construct a deep learning network designed to make predictions for energy and position for antineutrinos interacting in the detector. References for the various functions and their application programming specifics are available in \cite{pytorch} and \cite{tensorflow2015-whitepaper}. The simulated IBD signal samples correspond to the detector configuration, light yield, and calibration constants associated with the PROSPECT 2020 operational period ~\cite{2021}. Segment failures are included but detector response aging is not included. Selected random seed was 11 at the beginning of all runs, including the final 
training of the GAPE-selected models.

\begin{table} [!htbp]
\begin{center}
\begin{tabular}{|l||p{5cm}|}
\hline
\bf Optimizers & \bf Value  \\
\hline
RMSprop  & momentum=0.0 to .9 (increments of .1) \\ & epsilon= 1e-4 to 1e-8 (increments of factor 10) \\ & centered=False \\ 
\hline
Adam (ref included)  & beta 1=.80 to .99 (increments of .01) \\
& beta 2=.99 to .999 (increments of .001)\\
& epsilon= 1e-4 to 1e-8 (increments of factor 10) \\
\hline
Adadelta   & rho = .89 to .99 (increments of .02) \\
& epsilon = 1e-4 to 1e-8 (increments of factor 10) \\
\hline
Adagrad   & initial accumulator value= .05 to 0.3 increments of .05\\
& epsilon= 1e-4 to 1e-8 (increments of factor 10) \\
\hline
Nadam   & beta 1=.80 to .99 (increments of .01) \\
& beta 2=.99 to .999 (increments of .001)\\
& epsilon= 1e-4 to 1e-8 (increments of factor 10) \\
\hline
Adamax   & beta 1=.80 to .99 (increments of .01) \\
& beta 2=.99 to .999 (increments of .001)\\
& epsilon= 1e-4 to 1e-8 (increments of factor 10) \\
\hline
SGD & momentum=0.0 to .9 (increments of .1) \\ & Nesterov=False \\
& Dampening=0\\
\hline
\end{tabular}
\end{center}
\caption{Optimization functions available within gene pool along with the allowed ranges for parameters specific to each optimizer. These optimizers were selected for the gene pool based on their popularity amongst the machine learning community at the time of this writing. Epsilon is a small number that is used to avoid dividing by zero and thereby helping to maintain numerical stability. A number too small can lead to excessively large weights but a number too large can prevent efficient weight updates. Beta 1 and Beta 2 refer to the exponential decay rates for the 1st and 2nd momenta estimates. Note however that a Beta 1 or Beta 2 that is too large tends too make the learning slower. For the Adadelta and the RMSprop optimizer, the rho is a decay rate which is applied to the learning rate.  }
\label{optimizers}
\end{table}

\begin{table} [!htbp]
\begin{center}
\begin{tabular}{|p{6cm}|l|}
\hline
\bf Loss Functions: Regression  \\
\hline
MeanAbsoluteError \\

MeanAbsolutePercentageError \\

MeanSquaredError \\

CosineSimilarity \\

Huber \\
logcosh \\
\hline
\bf Loss Functions: Classification  (S)\\
\hline

Binary Cross-entropy \\
Poisson \\
Hinge Loss \\
Squared Hinge Loss \\

\hline
\bf Loss Functions: Classification  (M)\\
\hline

Cross-Entropy \\
Kullback Leibler Divergence Loss \\

\hline
\end{tabular}
\end{center}
\caption{Table of loss functions available within gene pool divided by target type. (S) stands for single target and (M) stands for multi-target.}
\label{loss}
\end{table}

\begin{table} [!htbp]
\begin{center}
\begin{tabular}{|l||p{5cm}|}
\hline
\bf Activation Functions & \bf Alpha \\
\hline\
relu  & 0.0 - 2.0 (.25 increments)\\
\hline
elu  & 0.0 - 2.0 (.25 increments)\\
\hline
selu  & alpha=1.67326324 (default) \\
\hline
tanh  & None\\
\hline
prelu  & alpha=0.0 (default)\\
\hline
leaky-relu  & 0.0 - 2.0 (.25 increments)\\
\hline
exponential  & None\\
\hline
\end{tabular}
\end{center}
\caption{Activation functions available within gene pool along with the allowed ranges for the alpha value which provides the slope for evaluations below threshold. The GA allows each layer to have a unique parameter for choosing an activation function. }
\label{activations}
\end{table}

\begin{table} [!htbp]
\begin{center}
\begin{tabular}{|l||p{4cm}|}
\hline
\bf Additional Instructions & \bf Value  \\
\hline
Number of Hidden Layers  & 0 - 5\\
\hline
For SOI Classifier and Energy Estimator: \\
Neurons per Hidden Layer  & 25 - 2000 (increments of 25)\\
For IBD Classifier: \\
Neurons per Hidden Layer  & 80 - 1440 (increments of 16)\\
\hline
Dropout Rate   & 0 to 10 (increments of 1)$\%$\\
\hline
Learning Rate  & .0001 to .1 (increments of 10)\\ 
\hline

Batch Size & 16, 32, 64, 100 \\
\hline
Kernel Constraints  & None\\
& Max Norm (max value=2, axis=0)\\
& Unit Normal \\

\hline
Initializer  & Glorot Normal\\
& Glorot Uniform \\
& Constant \\
& Identity \\
& Ones \\
& Orthogonal \\
& Zeros \\
& HeUniform (minval=0., maxval=1.)\\
& HeNormal (minval=0., maxval=1.)\\
& Random Normal (mean=0., std dev=1.) \\
& Random Uniform (minval=0., maxval=1.)\\
& Lecun Normal \\
& Lecun Uniform \\
& Normal \\
& Uniform \\
& Truncated Normal \\

\hline
\end{tabular}
\end{center}
\caption{Additional genes available within the pool along with the allowed ranges for parameters specific to each item. It was found that increments of less than approximately 20 neurons did not create statistically different predictions. Neurons were randomly removed at rates ranging between between 0 and 10 percent in order to better generalize the model. Note that each layer has its own unique dropout rate parameter that is provided by the GA. Initializers are chosen to initialize the weights in distinct variations that precede training and are chosen individually for each layer. Kernel constraint is another regularization technique that checks the size of the weights and rescales them if they surpass a certain threshold (constraints were not used on the bias weights and bias initializers were set to 'zeros') and these are also applied separately for each hidden layer.  }
\label{other}
\end{table}

\begin{table} [!htbp]
\begin{center}
\begin{tabular}{|p{6cm}|l|}
\hline
\bf Regularization of Features  \\
\hline
None \\

Standard Scaler \\

Min Max Scaler  \\

\hline
\end{tabular}
\end{center}
\caption{Scaling options for the features. The options are: the standard scaler which follows a standard normal distribution ($\mu$=0, and unit variance), a min max scaler which scales each feature individually to a given range (0 to 1 for positive values and -1 to 1 for features that contain negative values), and no feature scaling. In rare cases no feature scaling yields the best result which was the case for the energy estimator.}
\label{standards}
\end{table}

\newpage
\makeatletter
\let\appendix\originalJMLRappendix 
\renewcommand{\thesection}{\arabic{section}}
\setcounter{section}{0}
\makeatother

\vskip 0.2in
\bibliography{mybib}

\end{document}